\UseRawInputEncoding
\documentclass{sf2a-conf2021}
\usepackage{graphicx}
\usepackage{hyperref}
\usepackage[]{natbib}  
\usepackage{amssymb}
\usepackage{epstopdf}

\def\BibTeX{{\rm B\kern-.05em{\sc i\kern-.025em b}\kern-.08em
    T\kern-.1667em\lower.7ex\hbox{E}\kern-.125emX}}
\bibpunct{(}{)}{;}{a}{}{,}  

\begin{document}

\TitreGlobal{SF2A 2021}


\title{Exploring the Cosmic Dawn with NenuFAR}

\runningtitle{}

\author{F.G. Mertens$^{1,2}$}

\author{B. Semelin}\address{LERMA, Observatoire de Paris, PSL Research University, CNRS, Sorbonne Universit\'e, F-75014 Paris, France}

\author{L.V.E. Koopmans}\address{Kapteyn Astronomical Institute, University of Groningen, PO Box 800, 9700 AV Groningen, The Netherlands}

\setcounter{page}{237}


\maketitle


\begin{abstract}
The exploration of the Cosmic Dawn, the period of the Universe during which the first stars and galaxies were formed, is one of the last frontiers of modern astronomy and cosmology. The redshifted 21-cm line emission from neutral hydrogen is a unique probe and can open this era for astrophysical and cosmological studies. The tentative detection of the 21-cm global signal by the EDGES team at $z \sim 17$ underlines the need for an interferometric detection of this signal to discriminate between the numerous models trying to explain this unexpected discovery. The NenuFAR Cosmic-Dawn Key-Science Program aims to perform this detection in the redshift range $z \sim 15 - 31$ with a novel SKA precursor, the NenuFAR radio telescope located at the Station de Radioastronomie de Nançay and that started operating in 2019. Due to its compactness it is particularly sensitive to the large scale of the 21-cm signal. Only 100 hours of observation are needed to reach the level of the most extreme models, while 1000 hours are needed for the more standard models. Observations have already started, accumulating to almost 500 hours on the North Celestial Pole field. In this contribution, we introduce the project, our first results and the developments in calibration and RFI mitigation specific to this new instrument.
\end{abstract}



\section{Introduction}

The observation of the 21-cm line of neutral hydrogen is the most promising method to study the Cosmic Dawn and Epoch of Reionization (EoR) in detail. Its detection can inform us about the timing and mechanisms of the formation of the first stars, as well as the impact on the physics of the interstellar medium and the intergalactic medium (IGM) of the radiation emitted by these first sources of light. Many observational programs are underway to detect it. The so-called ``global'' experiments, such as EDGES~\footnote{Experiment to Detect the Global Epoch of Reionization Signature, \url{https://loco.lab.asu.edu/edges}}~\citep{Bowman18}, aim at detecting the 21-cm signal average in the sky. In addition to the global signal, spatial fluctuations in the brightness temperature of the 21-cm signal can be detected by radio-interferometric imaging, using interferometers such as NenuFAR~\footnote{New Extension in Nançay Upgrading LOFAR, \url{https://nenufar.obs-nancay.fr}}~\citep{Zarka20}, LOFAR~\footnote{Low-Frequency Array, \url{www.lofar.org}}~\citep{vanHaarlem13}, MWA~\footnote{Murchison Widefield Array, \url{www.mwatelescope.org}}~\citep{Tingay13}, HERA~\footnote{Hydrogen Epoch of Reionization Array, \url{https://reionization.org}}~\citep{DeBoer17}, and in a near future SKA~\footnote{Square Kilometre Array, \url{www.skatelescope.org}}~\citep{Koopmans15}. The power spectrum of the image cube thus obtained allows in particular to characterize the fluctuations of the spin temperature, of the hydrogen density, and of the ionization fraction, on scales ranging from a few arc minutes to a few degrees.

The tentative detection reported by the EDGES team of an absorption spectrum at redshift $z \sim 14 - 21$~\citep{Bowman18}, in the range where the Cosmic Dawn is predicted, has unexpected features. This observed absorption signal (-500 mK $\pm$ 200 mK, 99 \% IC) is considerably deeper than expected~\citep{Fraser18}. Standard models predict a maximum absorption of the order of -200~mK and, to explain this observation, a new process must be introduced, either by supercooling the Inter Galactic Medium (IGM) gas~\citep[][e.g.]{Barkana18} or by an additional source of background radiation more intense than the CMB~\citep[][e.g.]{Fialkov19}. A confirmation is still needed, that could come from the SARAS2 experiment~\citep{Singh17}, and the EDGES detection might just be an artefact~\citep{Singh19}, but if true, this detection would have dramatic implications for our understanding of the physical processes that occurred during the Cosmic Dawn.n

\begin{figure*}
    \center
    \vspace{-0.7em}
    \includegraphics[width=0.8\linewidth]{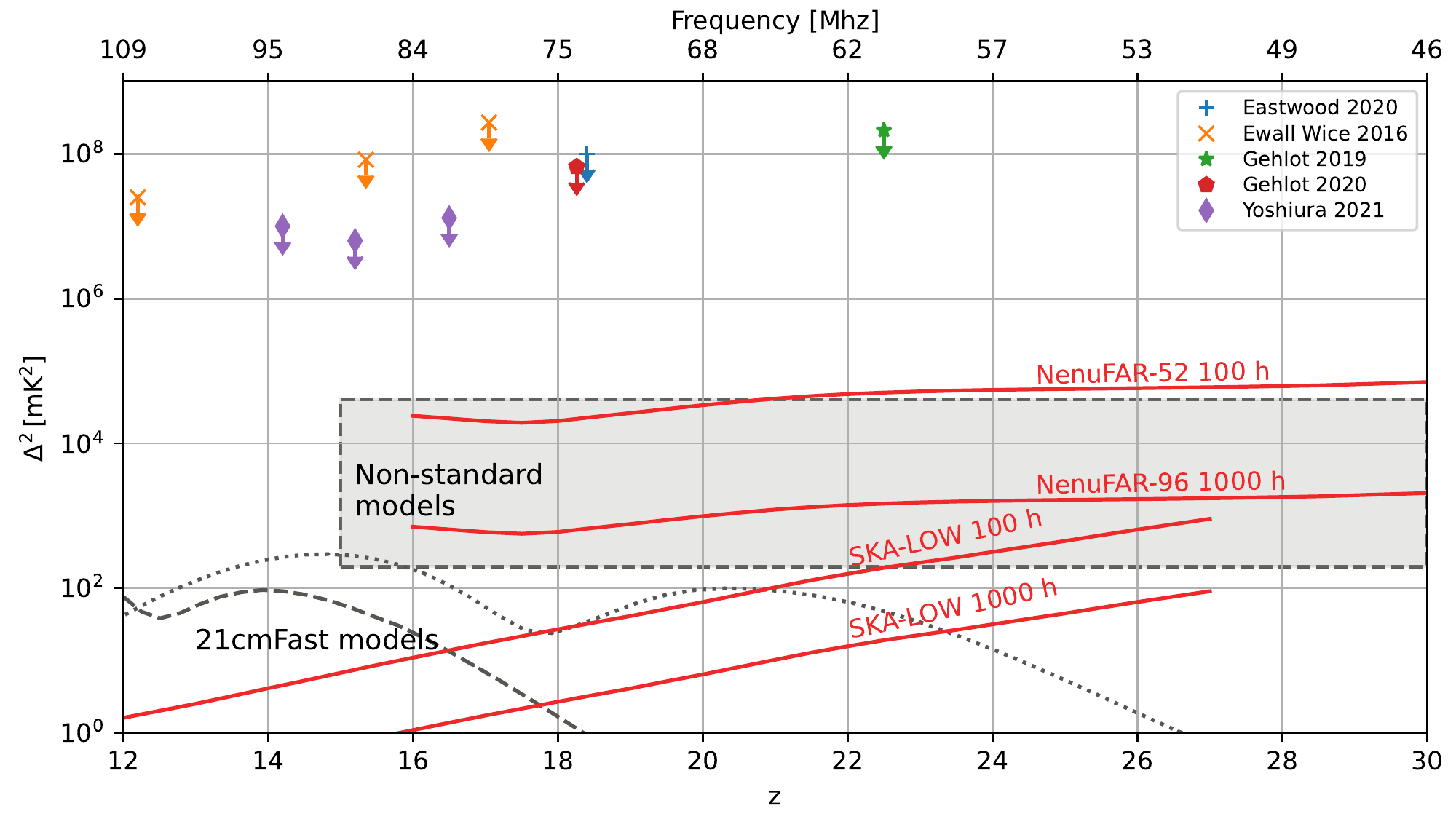}
    \vspace{-0.7em}
    \caption{Upper limits on the 21\,cm power spectrum at 95\% confidence ($2\sigma$) from various experiments from $12<z<30$, spanning the Cosmic Dawn, and at $k \gtrsim 0.1 {\rm Mpc}^{-1}$. Forecasted thermal noise limited $2\sigma$ sensitivity curves for the NenuFAR and SKA telescopes are also plotted for a 100 hour and 1000 hour integration}
    \vspace{-0.7em}
    \label{fig:upper_limit}
\end{figure*}

The global spectrum alone cannot distinguish between the different models explaining the signal observed by EDGES, an additional constraint is thus necessary. This may come from an interferometric detection, which will hold more information than the global signal. The different models also predict a power spectrum of the signal at 21-cm much more intense than the standard models predict which makes a detection much more accessible to the current instruments, and in first place by NenuFAR (see \autoref{fig:upper_limit}). We present here the effort by the NenuFAR Cosmic Dawn Key Project team to observe this signal with NenuFAR.

\section{The NenuFAR Cosmic Dawn Project}

NenuFar is a new radio telescope being installed at the Station de Radioastronomie in Nançay (France). In its final specifications, it will consist of 96 mini-arrays (MAs) included in a 400 m diameter disk and each composed of 19 dual polarized antennas. The compactness of the instrument, offering a large effective collecting area and short baselines, gives it a high sensitivity at large spatial scales, making it ideal to detect the 21-cm signal from the Cosmic Dawn. The NenuFAR Cosmic Dawn project is one of the early scientific Key Project (KP) of the NenuFAR telescope and aims at detecting this signal in the frequency range $40-85$ MHz ($z \sim 15-31$). This exciting goal is challenged by the difficulty of extracting the feeble 21-cm signal buried under astrophysical foregrounds orders of magnitude brighter and contaminated by numerous instrumental systematics. We estimate that $\sim $100h of observation are required to reduce the thermal noise to a level suggested by some non-standard models~\citep{Barkana18}, and $\sim$1000h for more standard models (see~\autoref{fig:upper_limit}). The project has been allocated more than 1000h of observations which started in September 2020 and will continue until early 2022. The target is the North Celestial Pole field (NCP) which is also one of the main target of the LOFAR-EoR project.

In these early stages of this new instrument, many components are still being installed or upgraded. Mid 2019, 52 MAs had been installed and new MAs are being added gradually. The remote MAs, which will provide when completed angular resolution down to 4 arcmin at 85 MHz, are also still in the process of being installed. Three of the 6 planned remote MAs are in operation at the time of writing. A new correlator (NICKEL) based on the COBALT2 LOFAR correlator~\citep{Pandey20}, increasing the instantaneous bandwidth and spectral resolution of the imaging mode, is also fully operational since mid-2020.
  
\section{Early Observation Phase}

In the first two semesters of the early science phase, before the availability of the NICKEL correlator and in preparation for later observations, we observed with NenuFAR a total of 340 hours on the North Celestial Pole field. The LaNewBa correlator available at that time offered only an instantaneous bandwidth of $\sim~3.1$ MHz and repeated observations were necessary to cover the full $30-85$ MHz bandwidth. Our goal in this first phase of observation was to build a Galactic and extragalactic model of the NCP field, assess potential technical issues and repeat observations to refine observing strategy for the second phase.

\begin{figure*}
    \vspace{-0.9em}
    \center
    \vspace{-0.8em}
    \includegraphics[width=0.8\linewidth]{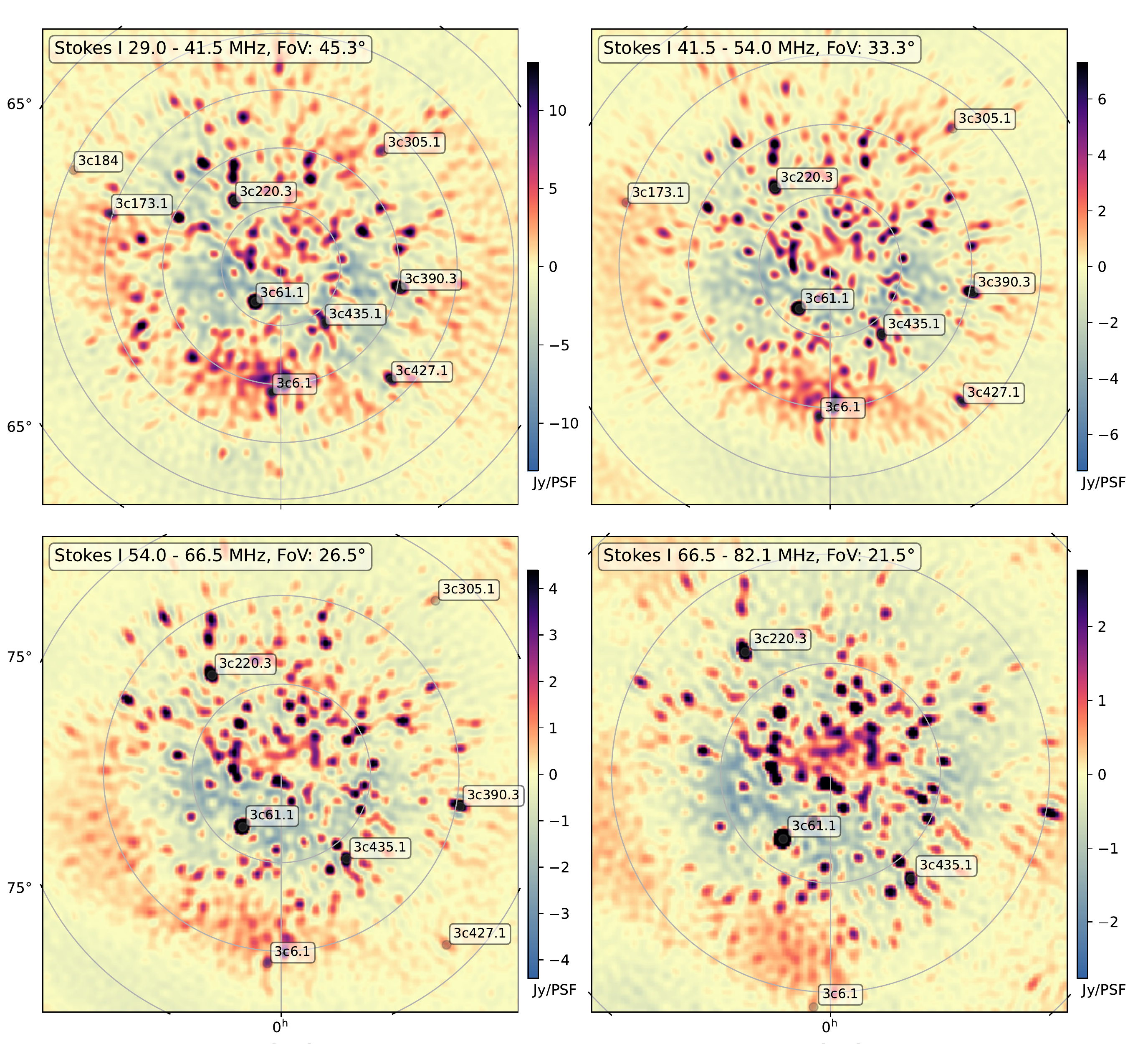}
    \caption{Wide-field Stokes I image of the North Celestial Pole field at four frequency ranges. The images are cleaned and produced using Briggs weighting. Integration time from top-left to bottom right is: 14h (rms noise: 161 mJy/PSF), 14h (rms noise: 80 mJy/PSF), 15h (rms noise: 56 mJy/PSF) and 12h (rms noise: 60 mJy/PSF). Both compact as well as Galactc diffuse emission are seen in these images.}
    \vspace{-0.7em}
    \label{fig:nenufar_ncp}
\end{figure*}

Radio interferometric observations are affected by numerous effects which need to be precisely corrected or accounted for to produce science ready data. After data transfer to the LOFAR-EoR cluster `Dawn' and pre-processing, the processing pipeline consists, in essence, of (1) Radio Frequency Interference (RFI) flagging with AOFlagger, (2) Calibration on three directions: NCP field, Cygnus A and Cassiopeia A, applying the solution of the NCP field and subtracting the contributions from Cygnus A and Cassiopeia A, (3) imaging using WSClean~\citep{Offringa14}, (4) power-spectra using the Lofar-EoR PS pipeline~\citep{Mertens20}. 

Emphasis was given in this first phase on data quality assessment. Besides looking at standard RFI flagging statistics, we also used the technique of near field imaging~\citep{Paciga11} to locate potential sources of local RFI in the field. Several were identified and reported with this technique.

The other aim of this first phase was to build a broadband Galactic and extragalactic sky model of the NCP field. All the observing nights of the NCP field have been processed and combined to form wide-field Stokes I and V images cubes. \autoref{fig:nenufar_ncp} shows wide-field Stokes I images of the North Celestial Pole field at four frequency ranges. In these images, the field around the NCP is confusion limited, but we also observe many 3C radio sources at distances up to 50 degrees from the NCP. At the highest frequency, we could detect, and cross-identify with catalogs, almost 200 sources inside the 20 degrees FoV of the beam-corrected image with an intrinsic flux higher than 2 Jy/PSF ($\sim$2 time the confusion noise standard deviation). The Galactic diffuse emission is also clearly visible at all frequencies and shows striking similarity with an image obtained at higher frequency (122 MHz) with the AARTFAAC system~\citep{Gehlot19}. We also produced Stokes I angular power spectra of the NCP field for all observed subbands. The galactic diffuse emission dominates on $\ell < 100$ (steep angular power spectrum index) while the confusion limited compact sources dominate the smaller scales (flat spectra). For the former, we observe an angular power spectrum index $\alpha \sim -2.5$ for the frequencies $> 50$ MHz which is consistent with previous observations of the NCP at other frequencies. The temperature spectral index is observed to be $\beta \sim -2.8$ for $\ell > 100$. 


\section{Phase 2: Observations and First Results}

During the second phase of observation, we aim to reach the 1000h of observations on the NCP field required for the detection of the Cosmic Dawn’s signal. A total of $\sim$500h have already been accumulated between July 2020 and June 2021 but which did not include yet the longer baselines. With the new NICKEL correlator narrowband RFI are now more optimally flagged which improved the data quality considerably. The availability of new remote additional MAs will also provide higher angular resolution required for an important milestone in the project which is to build a higher resolution and deeper sky model of the NCP field. This will then be used to improve our calibration, and allow for more accurate sky model subtraction with direction-dependent calibration. The first remote station is available since the beginning of October 2020 and commissioning observations have already been taken in this configuration.

Combining all our October observations, we produced a preliminary image cube at a resolution up to $\sim$15 arcsec. Artefacts are still present in these images due to direction-dependent effects and the yet-limited uv-coverage. The quality of these images will improve once later remote stations are integrated into the array. 

\section{Summary}

The NenuFAR Cosmic Dawn project aims at detecting the 21-cm signal from the Cosmic Dawn with the new NenuFAR telescope. The signal may be more intense than previously thought which may be the sign of exciting new science. In a first phase of observation programme, with limited capacity of the telescope, we build a preliminary sky model of the NCP field with a peculiar emphasis on assessing data quality, thus providing important feedback for operating the telescope. The second phase started in July 2020 and we aim to reach the 1000 hours of observations which may be required for the detection of the signal by early 2022.

\bibliographystyle{aa}  
\bibliography{mertens_S7} 

\end{document}